\documentclass[12pt]{article}

   \usepackage[round,authoryear]{natbib}

 \usepackage{ifpdf}
 \usepackage[pdftex]{epsfig}
 \usepackage[latin1]{inputenc}

 \usepackage{authblk} 

 \usepackage{amsfonts}
 \usepackage{amsthm}
  \usepackage{amssymb, latexsym, amsmath}
 \usepackage{amsbsy}
 \usepackage{amstext}
 \usepackage{amsgen}

 \usepackage{graphicx}
 \usepackage[normalem]{ulem}
 \usepackage[usenames, dvipsnames]{color}
 \usepackage{wasysym}
 \usepackage{verbatim}
 \usepackage{color}
 \usepackage[T1]{fontenc}
 \usepackage[scale=0.8]{geometry}
\usepackage{fancyhdr}
\usepackage{bm}

  \definecolor{mypink}{rgb}{0.858, 0.188, 0.478}
  \definecolor{mygrey}{rgb}{0.55, 0.68, 0.55}

 \definecolor{rkka}{RGB}{219,66,32}

 \newcommand{\s}{\nobreak\hspace{.11em}\nobreak}

 \newcommand{\be}{\begin{equation}}
 \newcommand{\ee}{\end{equation}}
 \newcommand{\ba}{\begin{eqnarray}}
 \newcommand{\ea}{\end{eqnarray}}
 \newcommand{\bs}{\begin{subequations}}
 \newcommand{\es}{\end{subequations}}

\begin{document}
  \title{
 {\Large{\textbf{{{Tidal quality of the hot Jupiter WASP-12b}
 ~\\~\\}
            }}}}

\author{

                                          {\Large{Michael Efroimsky}}\\
                                          {\small{US Naval Observatory, Washington DC 20392 USA}}\\
                                          {\small{michael.efroimsky$\,$@$\,$gmail.com
                                             }}
   \vspace{3mm}
   ~\\
 %
 {{and}}
 \vspace{4mm}
 ~\\

                                          {\Large{Valeri V. Makarov}}\\
                                          {\small{US Naval Observatory, Washington DC 20392 USA}}\\
                                          {\small{valeri.makarov$\,$@$\,$gmail.com
                                             }}
  }

     \date{}

 \maketitle

 \begin{abstract}
WASP-12b stands out among the planets of its class of hot Jupiters because of the observed fast orbital decay attributed to tidal dissipation. The measured rate of the orbital period change is
$\stackrel{\bf\centerdot}{\textstyle{P}}_{\rm orb}\,=\,-\,29\pm3\;\mbox{ms/yr}\;$=$\;(9.2\pm1.0)\times10^{-10}\;\mbox{s/s}$. In the literature heretofore, all attempts to explain this high rate were based on the assumption that the orbital evolution is dominated by the tides in the star. Since the modified tidal quality factor in yellow dwarfs is insufficient to warrant such a decay rate, a hypothesis was put forward that the star may actually be a subgiant. Using the latest data from the Gaia mission, we deduce that WASP-12 at $1.36\,M_{\sun}$ is an evolving dwarf at an early stage of post-turn-off evolution that has not yet depleted hydrogen in its core. Its unremarkable position in the color-magnitude diagram and the existence of close planets orbiting red giants of similar mass contradict the hypothesis of an abrupt boost of tidal quality due to structural internal changes. On the other hand, the previous research neglected the tidal dissipation in the planet assuming it to be negligible due to the likely synchronisation of its rotation and a presumed high quality factor. We critically reassess this assumption in the light of recent astrometric results for Jupiter and Saturn. Assuming that the structure of WASP-12b is similar to that of our Jupiter and Saturn, we find that the observed orbital decay is well explained by the tides in the planet. The estimated value of the planet's modified quality factor coincides almost precisely with that of our Jupiter.
 \end{abstract}

\section{Introduction}

An archetypal hot Jupiter exoplanet orbiting a solar-type star with a period of 1.09 d \citep
{2009ApJ...693.1920H}, WASP-12b has been for over a decade a testbed for theories of star-planet interactions, owing to its unusual properties. Even though not the closest exoplanet known to date \citep[with, e.g., WASP-19b's orbital period being only 0.7889 d, see ][]{2010ApJ...708..224H}, WASP-12b attracts much attention for its exceptionally high rate of orbital decay measured via transit time variations. This rate, $\stackrel{\bf\centerdot}{\textstyle{P}}_{\rm orb}=-29\pm3$ ms yr$^{-1}$, corresponds to a characteristic in-spiral time of 3.2 Myr \citep{2017AJ....154....4P}, cf. also the earlier result by \citet{2016A&A...588L...6M}. An even faster shrinkage of the orbit is obtained from observations with TESS
\citep[$ P_{\rm orb} / \stackrel{\bf\centerdot}{\textstyle{P}}_{\rm orb} =2.90 \pm 0.14 $ Myr,][]{Turner2021}. Doubts lingered for a while if the observed drift in transit times is caused by actual acceleration or by apsidal precession due to a third body. The latest considerations and evidence supports the former mechanism, as no sufficiently close companion has been found \citep
{bailey, 2020ApJ...888L...5Y}. If the orbital evolution is caused by the tidal dissipation of kinetic energy in the star, its modified quality factor should be a surprising $1.8\times 10^5$, which is at least two orders of magnitude lower (i.e., the tidal dissipation rate is two orders of magnitude higher) than what is expected from a solar-type MS dwarf. \citet{2018ApJ...869L..15M} discussed this puzzle and pointed out that, while a much higher dissipation can be theoretically expected from subgiant stars that have depleted their hydrogen fuel in the cores, the available evidence favored WASP-12A being an MS dwarf. Those authors speculated that an obliquity (latitude) tide in the planet could provide the missing power if the orbit is secularly perturbed by an outer planetary companion, so far undetected. Apart from the orbital decay, additional heating of the planet is required to explain the observed bloated radii of hot Jupiters in general and of WASP-12b in particular \citep{2010ApJ...713..751I}, leading to a degeneracy of model fitting for eccentricity, quality factor, and the mass of a possible solid core. Observations of the secondary eclipse of WASP-12b in the NIR joined the fray of conflicting signals, with \citet{2011AJ....141...30C} pointing out that the 0.5 phase of the secondary eclipse rules out a measurable eccentricity, a conclusion also supported by \citet{2011ApJ...727..125C}. Spectroscopic radial velocity measurements, on the other hand, clearly show the signs of a finite eccentricity of 0.049 \citep{2011MNRAS.413.2500H}.

Employing a phenomenological approach, we examine in this Letter which points of distinction can account for the features of the WASP-12 system. One possibility discussed in the literature is that the tidal dissipation in the host star is uniquely effective because it is in a short-lived subgiant stage of evolution. The other specific feature is the presence of distant stellar components making it a quadruple system. The stellar companion $1.06$ arcsec away from the primary was discovered by \citet
{2013MNRAS.428..182B} from the ground and confirmed by \citet
{2012ApJ...760..140C} with the HST. \citet
{2014ApJ...788....2B} further resolved this companion into a pair of twin M dwarfs separated by $84.3\pm 0.6$ mas (corresponding to 34.8 AU at the distance of WASP-12). Although the separation between this pair and the primary is at least 438 AU, \citet
{2017MNRAS.470.1657H} suggested that the inner orbit's eccentricity can be secularly excited in hierarchical quadruple systems like WASP-12 via the Lidov-Kozai perturbation mechanism. The outer perturber of WASP-12b being a binary system gives rise to additional dynamical effects via the possibly chaotic long-term evolution of the triple stellar system \citep{2017MNRAS.466.4107H}, although the short-term effect on the planet is probably insignificant.

 \section{Classification of the star}

  The properties of the system WASP-12 and the decaying orbit of its hot Jupiter have been addressed in numerous publications; see e.g. \citet{chernov} for an early comparison of observations and theory.
  At some point, the nature of the star became subject to debate. \citet{2009ApJ...693.1920H} refer to it as a `{\it{late-F star evolving off the zero age main sequence.}}' While most authors (e.g., \citeauthor{Turner2021} \citeyear{Turner2021}) agree that the star is F-type, \citet{weinberg} hypothesised that it may be a subgiant.
  Using the estimated effective temperature and mean density with their substantial uncertainties, \citet{weinberg} concluded that the properties of WASP-12A may be ambiguous with respect to its evolutionary status, mass, and age.
  The motivation for the hypothesis came from the theoretical surmise that the tidal dissipation rate for subgiants of this size is at least two orders of magnitude higher than that of MS dwarfs of similar mass, and that it dominates the orbital decay of its planet. Below we shall question this assumption. A detailed modeling by \citet{bailey}
  demonstrated that, while no model seems to fit all of the observational constraints on the star, `{\it{main-sequence models are less discrepant than subgiant ones}}.'

\begin{figure}[htbp]
\includegraphics[width=6.47cm]{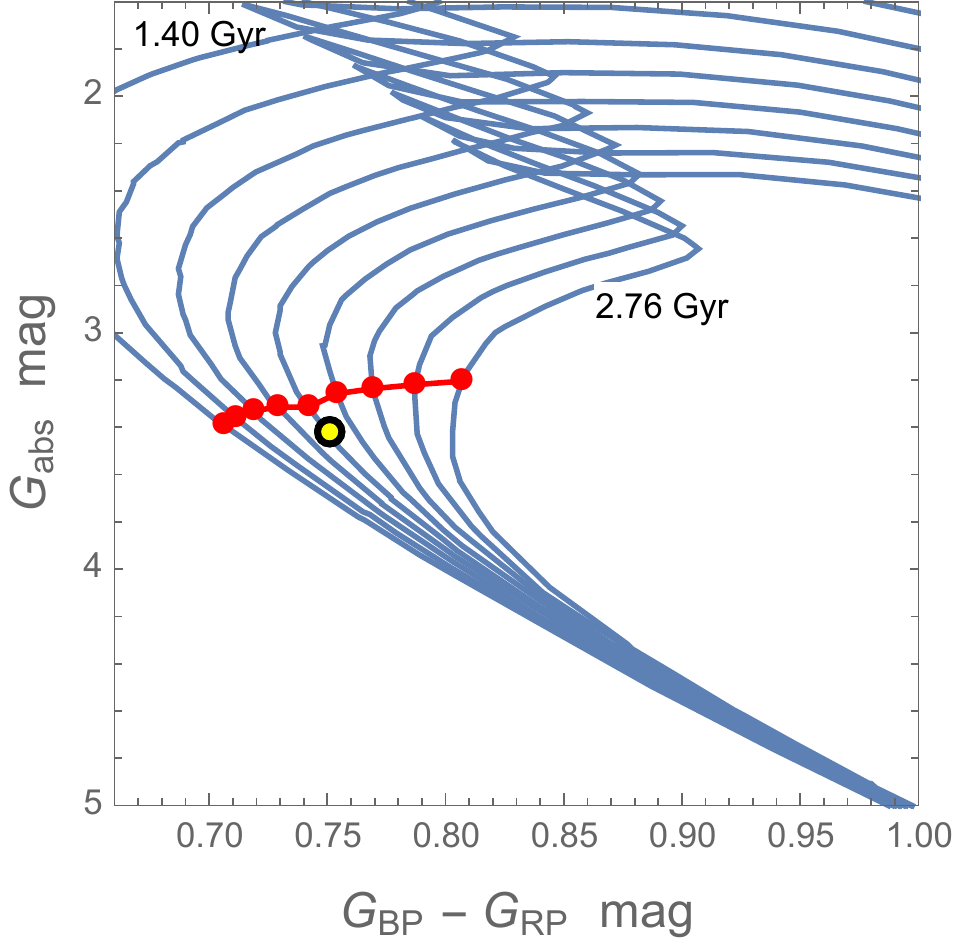}
\caption{Isochrones from the PARSEC suite of stellar models for 9 equally spaced stellar ages
as labeled in the plot computed for the Gaia EDR3 photometric system. The observed location of WASP-12 is shown with an filled circle. The red line with dots shows the evolutionary track of the closest nodal initial mass 1.4 $M_{\sun}$ after the turn-off point. \label{iso.fig}}
\end{figure}

  Using state of the art astrometric and photometric data from Gaia EDR3 \citep
  {2016A&A...595A...1G, 2021A&A...649A...1G}, we place WASP-12A directly on the color-magnitude in the $G_{\rm abs}$ versus $G_{\rm BP}-G_{\rm RP}$ axes thus avoiding numerous model-dependent assumptions and uncertainties associated with derived physical parameters. Its location is shown with an open circle in Fig. \ref
  {iso.fig}. We note that the parallax for this star is substantially smaller than what was assumed by many authors, and the star is definitely overluminous by $\sim 0.3$ mag compared to MS dwarfs of its color. Using the PARSEC \citep
  {2012MNRAS.427..127B, 2014MNRAS.445.4287T} stellar evolution models for metallicity $Z=0.026$
  and a star-specific interstellar extinction of $A_V=0.26$ mag from \citet
  {2018MNRAS.475.1121G}, nine isochrones are computed at ages 1.40  through
  2.76 Gyr with a step of 0.17 Gyr, represented with solid curves. The assumed super-solar metallicity [M/H]=0.1 is within the range of the determinations collected in the PASTEL catalog \citep{2016A&A...591A.118S} and close to the determination by \citet{2012ApJ...757..161T}. The PARSEC stellar models in the Gaia photometric system has been validated on open clusters including the Hyades \citep
  [e.g.,][]{2021A&A...649A...6G}. The best-matching isochrone node is that for age 2.08 Gyr, mass 1.36 $M_{\sun}$, $\log\,T_{\rm eff}=3.80$ K, $\log\,g=4.146$.
These values are quite close to the stellar properties collected by \citet
  {2012MNRAS.426.1291S}. While the photometry and astrometry errors from Gaia EDR3 are negligible, the greatest uncertainty is associated with the imprecisely known metal content. If the true [Fe/H] is higher than 0.1, the isochrones shift to the red and the star is yonger and closer to the main sequence.

The red line with dots intersecting the isochrones shows the approximate evolutionary track of a star with an initial mass 1.4 $M_{\sun}$. We note that WASP-12A appears to be at the initial stage of this post-turn-off track, and it took at least 680 Myr after its departure from the MS to get to the present state. In a longer time interval ($\sim 1$ Gyr), the star will reach the first folding point marking the beginning of overall contraction due to exhaustion of hydrogen in the core, which will be a greatly faster process. We conclude that WASP-12A occupies an unremarkable position at an early stage of post-MS evolution, still burning hydrogen in its core, and still a 1 Gyr away from the subgiant phase characterised by significant structural changes.

 \pagebreak

\section{Tidal parameters of the WASP-12 system\label{1}}

  Throughout this paper, the parameters of the primary lack a prime: $M$, $R$, $k_{l}$, $\epsilon_l$, $Q_{l}$, while those of the secondary are primed: $M^{\,\prime}$, $R^{\,\prime}$, $k^{\,\prime}_{l}$, $\epsilon^{\,\prime}_l$,  $Q^{\,\prime}_{l}$, etc. Conventionally, we omit the subscript of the quadrupole quality factors: $Q\equiv Q_2$ and $Q^{\,\prime}\equiv Q^{\,\prime}_2$.

  The Love numbers $k_l$, phase lags $\epsilon_l$, and quality factors $Q_l\equiv|\sin\epsilon_l|^{-1}$ are functions of the Fourier tidal modes $\omega_{lmpq}$.
    \footnote{~We remind that the tidal modes in a perturbed body are given by \citep{EfroimskyMakarov2013}
 \ba
 \omega_{lmpq}\;\equiv\;(l-2p)\;\dot{\omega}\,+\,(l-
 2p+q)\;n\,+\,m\;(\dot{\Omega}\,-\,\dot{\theta})\,~.
 \nonumber
 \ea
 where $lmpq$ are integers, $\omega$ and $\Omega$ are the argument of the pericentre and the longitude of the node, $\dot{\theta}$ is the rotation rate of the body, while $n\s\equiv\s{\bf{\dot{\cal{M}}}}$ is the mean motion, $\s{\cal{M}}\s$ being the mean anomaly.  As can be observed from equation (11) in {\it{Ibid.}}, the physical forcing frequencies are equal to these modes' absolute values:
 $
 \chi_{lmpq}~\equiv~|\,\omega_{lmpq}\,|~.
 $
 }
  The {\it{quality functions}} are \citep[cf. ][]
  {2018ApJ...857..142M}
  \ba
  K_l(\omega_{lmpq})\,\equiv\,k_l(\omega_{lmpq})\,\sin\epsilon_l(\omega_{lmpq})\,=\,\frac{k_l(\omega_{lmpq})}{Q_l(\omega_{lmpq})}\;\mbox{Sign}\,\omega_{lmpq}\;\,,
  \label{1}
  \ea
  \ba
  K_l^{\,\prime}(\omega^{\,\prime}_{lmpq})\,\equiv\,k^{\,\prime}_l(\omega^{\,\prime}_{lmpq})\,\sin\epsilon^{\,\prime}_l(\omega^{\,\prime}_{lmpq})\,=\,\frac{k^{\,\prime}_l(\omega^{\,\prime}_{lmpq})}{Q^{\,\prime}_l(\omega^{\,\prime}_{lmpq})}\;\mbox{Sign}\,\omega^{\,\prime}_{lmpq}\;\,.
  \label{2}
  \ea
   Another term sometimes used for them in the literature is {\it{kvalitet}}, a Danish word.

  The notation $Q^{\,\prime}$ reserved for the secondary's quadrupole quality factor, we equip the so-called {\it{modified quality factors}} with a dagger:
  \ba
  Q^{\dagger}\,\equiv\,\frac{3}{2}\;\frac{1}{|K_2|}\;\;,\qquad {{Q^{\,\prime}}}^{\,\dagger}\,\equiv\,\frac{3}{2}\;\frac{1}{|K^{\,\prime}_2|}\;\,.
  \label{3}
  \ea
  These quantities should not be confused with the actual quality factors $Q$ and $Q^{\,\prime}$.

  \subsection{Tidal dissipation in the star\label{1.1}}

 Slightly hotter than the Sun, WASP-12A is similar to it in mass and size:
 \ba
 M\,=\,1.36\,M_{\sun}\,=\,2.70\times 10^{30}\;\mbox{kg}\,,\qquad R\,=\,1.57\,R_{\sun}\,=\,1.09\times 10^9\;\mbox{m}\;,
 \label{size}
 \ea
 {though it may have a very different interior structure, being of the F type.}
 %
 According to \citet{barker}, dissipation in
 {slowly-rotating} F-type stars is dominated by internal gravity waves in radiation zones. For masses in the range of $1.2 - 1.4\, {M_{\sun}}$, the modified quality factor scales as \footnote{~Our formula (\ref{interpolation}) interpolates Barker's results for $1.2 \, {M_{\sun}} \leq M \leq 1.4\, {M_{\sun}}$, see the paragraph after equation (54) in \citet{barker}. Extension of this dependency down to $M={M_{\sun}}$ renders the value of $Q^\dagger = 10^4$ agreeing perfectly with the corresponding value from \citet{ivanov}, see the lower curves in their Figure 15.}
 \ba
 \log Q^{\dagger}\,=\;7\,+\,\left(\frac{M}{M_{\sun}}\,-\,1.3 \right)\times 10\,\;.
 \label{interpolation}
 \ea
 For WASP-12A, this entails $Q^{\dagger} =10^{7.6} = 3.98\times 10^7\,$, wherefrom
 \ba
 |K_2|\,=\,3.77 \times 10^{-8}
 \,\;.
 \label{}
 \ea

  \subsection{Tidal dissipation in the planet\label{1.2}}

 For the solar-system giant planets, astrometric measurements give \citep{Lainey2009,Lainey2017}:
 \ba
 |K^{\,\prime}_2| = (1.02 \pm 0.203 ) \times 10^{-5}\;\;,\quad\mbox{for Jupiter}\;\,,
 \label{Jupiter}
 \ea
 \bs
 \ba
 |K^{\,\prime}_2| = (1.59 \pm 0.74) \times 10^{-4}\;\;,\quad\mbox{for Saturn}\;\,,
 \label{}
 \ea
 high values interpreted by \citep{remus2015} as an argument in favour of the presence of a noticeable solid core, see also \citet{2010ApJ...713..751I}. Obtained from astrometric observations of the eight main and four coorbital moons, Saturn's $|K^{\,\prime}_2|$ demonstrated no obvious variations for tidal
 frequencies corresponding to those of Enceladus and Dione. A theoretical model developed by \citet{remus2012} explains this by attributing the values of
 $10^{14}$ - $10^{16}\,$ Pa s\, to the viscosity of the solid core.

 The hypothetical existence of a solid core does not necessarily diminish the importance of hydrodynamics. A dissipation peak,
 \ba
 |K^{\,\prime}_2| = (12.394 \pm 1.727) \times 10^{-4}\;\;,\quad\mbox{for Saturn, at Rhea's frequency}\;\,,
 \label{}
 \ea
 \es
 indicates the existence of an additional, frequency-dependent friction mechanism. While intense tidal damping in some of the moons cannot be excluded,
 another possibility could be turbulent friction acting on tidal waves in the fluid envelope of the planet \citep{Ogilvie}.\,\footnote{~
 {Several studies advocate models alternative to that containing a solid core. Recent Juno observations \citep{Wahl} indicate that the heavy elements are likely to be distributed  throughout a large portion of the interior (possibly, out to 0.5 to 0.6 of the planetary radius), and a core considered in the old interior model may not exist. Other mechanisms capable of producing high dissipation are gravity waves in stably-stratified layers of the planet \citep{Fuller,Andre}, and inertial waves in the convective envelope \citep{Ogilvie}. The idea of resonance locking of these two types of modes has also been proposed \citep{Fuller}.}
 }

 Assuming that the internal structure of Jupiters is universal,\,\footnote{~An additional argument in favour of the universality of Jupiters' structure comes from the recent observations of the close-in Jupiter WASP-103b whose quadrupole Love number turned out to be close to that of our Jupiter \citep{barros}.} we accept that the value of $|K^{\,\prime}_2|$ of WASP-12b is residing in the range of $10^{-5}$ -- $10^{-3}$.

 \section{Tidal rate of the semimajor axis\label{Section4}}

  A general formalism to describe the orbital evolution of a tidally perturbed two-body problem was suggested by Kaula (1964) whose work was an impressive mathematical extension of the ideas of Darwin (1880).\,\footnote{~In modern notation, Darwin's approach is explained in \citet{Ferraz}. For an introduction to the Kaula theory, see \citet{EfroimskyMakarov2013}.}  Both these classical works were based on a Fourier decomposition of the additional potential generated by tidal deformation. While Darwin explored the first several terms of the series, Kaula wrote down the full expansion. Also, while Darwin's approach implied a particular viscoelastic response of the body's material, Kaula's formalism was more adaptable and could, in principle, be combined with an arbitrary rheology.
  The main results of both Darwin's and Kaula's works were the expressions for the tidal rates of the Keplerian orbital elements. An accurate re-derivation from first principles, carried out by \citet{Boue}, pointed out several oversights in the old theory, especially in the expression for $di/dt$.

  From the general formulae for $da/dt$ provided in \citet{Boue}, it is possible to single out the $e^2$-approximations for the inputs in $da/dt$ generated by the tides in the partners. The quadrupole contribution from a synchronised secondary is
 \ba
 \left(\frac{da}{dt}\right)_{l=2}^{(sec)}=\;
 -\;57\;a\,n\,e^2\,\left(\frac{R^{\,\prime}}{a}\right)^{\textstyle{^{5}}}\frac{\;M\;}{M^{\,\prime}}\;K_2^{\,\prime}(n)\;+\;O({i^{\,\prime\,}}^2)\;+\;O(e^4)
 \,\;,
 \label{49}
 \ea
  while the contribution from a nonsynchronous primary is
   \ba
 \nonumber
 \left(\frac{da}{dt}\right)^{(prim)}= &-& 3\;n\,a\;\frac{M^{\,\prime}}{M\;}\,\left(\frac{R}{a}\right)^{\textstyle{^5}}\,K_{2}(2n-2\dot{\theta})\;
 ~\\
 \nonumber\\
 \nonumber
 &-& \frac{3}{8}\s n\s a\s e^2\s\frac{\;\,M^{\,\prime}}{M}\s\left(\frac{R}{a}\right)^{\textstyle{^{5}}}\s\left[
 -\s 40\, K_{2}(2n-2\dot{\theta})\s +\s 6\s K_2(n)\s +\s K_2(n-2\dot{\theta})\s +\s 147\s K_2(3n-2\dot{\theta})
 \s\right]
 ~\\
 \nonumber\\
 &-& \frac{3}{4}\;a\;n\,\left(\frac{R}{a}\right)^7\,\frac{M^{\,\prime}}{M}\;
 \left[\,5\;K_3(3n-3\dot{\theta})\;+\;K_3(n-\dot{\theta})\,\right]\;+\;O(i^2)\;+\;O(e^4)
 ~~_{\textstyle{_{\textstyle .}}}
  \label{comparison}
 \ea
 Equation (\ref{comparison}) contains also degree-3 terms of order $e^0$ because, owing to the extreme tightness of WASP-12, they are comparable to the quadrupole $e^2$ terms. At the same time, the degree-3 input $\;-\,\frac{\textstyle 3}{\textstyle 4}\,a\s n\s \left(\frac{\textstyle R^{\,\prime}}{\textstyle a}\right)^7\,\frac{\textstyle M\;}{\textstyle M^{\,\prime}}\, \left[\s 5\s K_3^{\,\prime}(3n-3\dot{\theta}^{\,\prime}) + K_3^{\,\prime}(n-\dot{\theta}^{\,\prime})\s\right]\;$ is omitted in expression (\ref{49}), because it
  vanishes for a synchronised secondary.
  To determine which body defines the tidal decay, we compare the leading contributions, i.e. divide expression (\ref{49}) by the first term from expression (\ref{comparison}):
    \footnote{~We used ${\textstyle R^{\,\prime}}/{\textstyle R} =
    \s 1.15\times 10^{-1}\s$ and ${\textstyle M}/{\textstyle M^{\,\prime}}
    \s=\s 9.68\times 10^2$.
    }
 \ba
 \nonumber
 \frac{\mbox{leading term, due to synchronised secondary}}{\mbox{leading term, due to nonsynchronous primary}}&=&19\,e^2\,\left(\frac{R^{\,\prime}}{R}\right)^5\left(\frac{M}{M^{\,\prime}}\right)^2
 \frac{K^{\,\prime}_{2}(n)}{K_{2}(2n-2\dot{\theta})}
 \label{}\;\qquad\\
 \nonumber\\
 &=&358\,e^2\;\frac{K^{\,\prime}_{2}(n)}{K_{2}(2n-2\dot{\theta})}\;\,.
 \ea
 In Sections \ref{1.1} and \ref{1.2}, we saw that $|K_2|\simeq 3.77 \times 10^{-8}$, while $|K^{\,\prime}_2|=10^{-5}$ -- $10^{-3}$. Consequently,
 \ba
  \frac{\mbox{leading term, due to synchronised secondary}}{\mbox{leading term, due to nonsynchronous primary}}\,=\;950\;e^2\times\left(\,10^{2}\;-\;10^{4}\,\right)\;\,.
 \label{}
 \ea
 Even at the lowest boundary of the Jupiter's dissipation, the dissipation in it is dominant for $e > 3.2\times 10^{-3}$. However, at the highest boundary of Jupiter's dissipation, the dissipation in it is dominant for $e>3.2\times 10^{-4}$.

 For $e=4\times 10^{-2}$, the dissipation in the planet is leading overwhelmingly:
 \ba
 \frac{\mbox{leading term, due to synchronised secondary}}{\mbox{leading term, due to nonsynchronous primary}}\,=\;1.52\times \left(\,10^{2}\;-\;10^{4}\,\right)\;\,.
 \label{}
 \ea
 The eccentricity of WASP-12b estimated directly from the second harmonic of the observed radial velocity curve ranges between 0.03 and 0.05. The fitted nodal longitude of the periastron, however, is close to $270^\circ$, which led \citet{2020ApJ...889...54M} to suggest an alternative explanation for the observed second harmonic. A semidiurnal variation of radial velocity can theoretically be caused by the tidal bulge on the tidally deformed star and the resulting variation of the projected radius and the surface velocity flow \citep{2012MNRAS.422.1761A}. The exposition of the latter paper is mostly based on the study by \citet{1977AcA....27..203D}, which suffers from a few omissions and ill-justified simplifications. In particular, the assumptions of incompressible fluid motion and the omission of the inertia terms in the employed hydrostatic model are likely to overestimate the magnitude of the induced motion, so that the estimated model uncertainty \citep[a factor of 2,][]{2012MNRAS.422.1761A} is lopsided. Inertial flows outside of spin-orbit resonances and compressibility are bound to decrease the amplitude of tidal deformation. This point is illustrated by the case of WASP-18Ab system, which is a close analog of WASP-12. Arras et al. (2012) specifically predicted a semidiurnal tidal signal with an amplitude of 32 m s$^{-1}$, which should be easily detectable. This would be seen as a false eccentricity of 0.0176 and $\omega_*=3/2\pi$. No such signal is seen in the data for WASP-18Ab. On the contrary, the most recent observational estimate by \citet{2020A&A...636A..98C} is $e=0.0051$.
 {Finally, we would mention that \citet{Bunting}, who employed a compressible and non-adiabiatic linearised numerical approach, found significant differences with \citet{2012MNRAS.422.1761A}. Their best-fitting model that gives the closest prediction for WASP-12 in terms of the amplitude of the RV signal, however, also predicts a large phase lag of the tidal deformation, which is not observed.}

{To conclude, even though the value $e\simeq0.04$ originally produced by the RV analysis may be interpreted as a confusion signal caused by the tidal deformation of the host star \citep[Section 5.3]{Bunting}, the theoretical basis is far from clear and solid. Further progress in this direction requires additional ways to constrain the orbital eccentricity. Reprocessing the available and obtaining new photometric information may, in principle, be useful to this end. A phase shift of the occultation time with respect to the transit time could be one of the manifestations of a nonzero eccentricity. Using the results obtained from TESS light curves by \citet{Turner2021}, we estimated for an adjacent pair of transit and eclipse a center time difference of $0.54541(107)$ d, which is well within $1\sigma$ of $P_{\rm orb}/2$. This is consistent with the previously RV-estimated longitude of periastron within one observational error of $270$ deg. The estimated average depth of secondary eclipse $480$ ppm is quite low for this photometrically noisy star, which probably explains the absence of eclipse duration estimates. Furthermore, independently fitting three eclipses of WASP-12b from ground-based observations, \citet{2019A&A...628A.115V} detected a strongly variable depth with a range of 1200 ppm and one formally negative value. Of course, they did not even try to estimate the eclipse duration simply assuming the eclipse to be a scaled copy of the mean transit. A large number of additional high-cadence light curves with the best instruments is needed to realize the possibility of placing contraints on the eccentricity.}

 \section{The rate $\stackrel{\bf\centerdot}{P\,}$ of the orbital period}

 Starting with $P_{\rm orb}=2\pi/n$, and approximating the anomalistic mean motion $n\equiv\dot{\cal{M}}$ with its osculating counterpart $\sqrt{G(M+M^{\,\prime})\,a^{-3}\s}$, we get $\s P_{\rm orb}\propto a^{3/2}\s$ and therefore $\s\frac{\textstyle \stackrel{\bf\centerdot}{P\,}_{\rm orb}}{\textstyle P_{\rm orb}}\s=\s\frac{\textstyle 3}{\textstyle 2}\s\frac{\textstyle \dot{a}}{\textstyle a}\s$. \,Thence,
 \ba
 \stackrel{\bf\centerdot}{P\,}_{\rm orb}\;=\,\frac{3}{2}\,\frac{P_{\rm orb}}{a}\;\dot{a}\,=\,\frac{3\,\pi}{n\,a}\,\dot{a}\;\;.
 \label{}
 \ea
 If we take into account only the leading input (\ref{49}) due to the synchronised secondary, the resulting approximation for the orbital period rate will read:
 \footnote{~We used: ${\textstyle R^{\,\prime}}/{\textstyle a} =
        \s 3.79 \times 10^{-2}\s$
        and ${\textstyle M}/{\textstyle M^{\,\prime}}
    \s=\s 9.68\times 10^2$.
    }
 \bs
 \ba
 \stackrel{\bf\centerdot}{P\,}_{\rm orb}&=&-\;171\;\pi\;\frac{M}{\;M^{\,\prime}}\,
 \left(\frac{R^{\,\prime}}{a}\right)^{\textstyle{^{5}}}e^2\,
 K^{\,\prime}_{2}(n)\;+\;O(i^2)\;+\;O(e^4)
 \label{porb.eq}\\
 \nonumber\\
 &=&-\;4.06\times 10^{-2}\;e^2\,K^{\,\prime}_{2}(n)\;+\;O(i^2)\;+\;O(e^4)
  ~~_{\textstyle{_{\textstyle ,}}}\;
 \label{our}
 \ea
 \label{}
 \es
 wherefrom
 \ba
  K^{\,\prime}_{2}(n)\;=\;-\;\frac{2.46}{e^2}\times 10^1\stackrel{\bf\centerdot}{P\,}_{\rm orb}
 \label{}
 \ea
 The eccentricity of WASP-12b is $e=0.04$, while the observed rate of the orbital period is $\stackrel{\bf\centerdot}{\textstyle{P}}_{\rm orb}\, = - \,29 \pm 3$ ms/yr = $-\,(9.2   \pm  1.0)\times 10^{-10}$ s/s. The insertion of these numbers entails:
 \ba
 K^{\,\prime}_{2}(n)\,=\,1.42\times 10^{-5}\quad\mbox{and}\quad {Q^{\,\prime}}^{\,\dagger}=\frac{3}{2}\,\frac{1}{|K^{\,\prime}_2|}\,=\,1.06\times 10^{5}\,,\quad
 \label{}
 \ea
 values different only by a factor less than $1.4$ from those measured for our own Jupiter, see equation (\ref{Jupiter}).

This estimation assumes that the observed asymmetry of the radial velocity curve is caused entirely by orbital eccentricity. How could a planet so close to the host star showing a remarkably fast orbital decay maintain a finite eccentricity? For a planet-dominated tidal dissipation scenario, the characteristic time of circularisation is much shorter than that of day shortening \citep{2018ApJ...857..142M}. A few speculative scenarios can be considered. A hitherto undetected external companion can excite a substantial obliquity of the planet's equator accounting for the increased tidal dissipation \citep{2018ApJ...869L..15M}. However, a planetary mass companion can also excite planet b's eccentricity to the observed value, and, as we show here, a large obliquity is no longer required. Another hypothetical possibility is an inner planet in a 1:2 mean motion resonance (MMR) that recently merged into the star. As long as it exists, the inner planet can keep the outer planet's eccentricity at a significant level depending on its mass \citep{2012ApJ...761...83M}. Finally, the outer layer of WASP-12b is bloated and inviscid enough to be locked in {\it pseudosynchronous} rotation \citep{2015ApJ...810...12M} causing additional friction with the synchronised rigid core. The orbital dynamics of such planets is yet to be explored.

 \section{Discussion}

 We have demonstrated that at the lowest boundary of Jupiter's dissipation, the orbital decay of WASP-12 is dominated by the tides in the planet for $e > 3.2\times 10^{-3}$; while at the highest boundary of Jupiter's dissipation, the decay is dominated by the tides in the planet for $e>3.2\times 10^{-4}$.
 For the observed value of $e=0.04$, we have calculated the value of the planet's tidal quality $|K_2^{\,\prime}|=\frac{\textstyle 3}{\textstyle 2}\,\frac{\textstyle 1\;}{\textstyle{Q^{\,\prime}}^{\,\dagger}}\,$, \,which happened to be only slightly higher than that of our Jupiter (in the case that this mechanism is the only contributor to orbital evolution).

 Despite the achieved progress, a more detailed modeling of this system is required. Specifically, it remains to be explored if the pull of the two red-dwarf companions is sufficient to create episodes with the Jupiter's eccentricity ascending to as high values as the observed $e=0.04$. At this point, it is unclear what other processes could be responsible for such high an eccentricity of such a close-in planet. One theoretical possibility is that the planet is not synchronous but is rotating significantly faster than the mean orbital motion, which would lead to a higher rate of tidal heating and boosting the orbital eccentricity. The latest observational evidence about a shift of the peak of planet's light curve by $0.049\pm 0.015$ toward the evening terminator \citep
 {2021MNRAS.503L..38O} gives some credibility to this scenario. Another open question is the  proximity of the obtained tidal quality of WASP-12b to that of our Jupiter. That $|K_2^{\,\prime}|=\frac{\textstyle 3}{\textstyle 2}\,\frac{\textstyle 1\;}{\textstyle{Q^{\,\prime}}^{\,\dagger}}\,$ of WASP-12b is higher is not surprising, given the higher temperature of WASP-12b.
 Surprising is the difference being so small.

 Given the short time of orbit circularisation by planet-dominated tides (in the case of synchronous rotation), the main difficulty of the proposed interpretation is how to achieve or maintain an eccentricity above this threshold. This seems to be impossible without the aid of an external agent, whether an undetected outer planet or a close inner planet in an MMR that recently merged with the host star.
 {Although exoplanet systems in exact 2:1 MMR are relatively rare \citep[Kepler-88 being the best known example discovered via the strong transit time variations of the inner planet,][]{2013ApJ...777....3N}, about 16\% of known transiting multiplanet systems have period ratios in the vicinity of this resonance \citep{2011ApJS..197....8L}. Such systems may arise from perfect resonance configurations due to the asymmetric effect of tidal dissipation as the planets get closer to the host star. Indeed, as follows from Eq. \ref{porb.eq}, the rate of period decline id proportional to $(R'/a)^5$, and the inner planet of the same radius would spiral in 10 times faster than its outer companion, given all other parameters the same. The observed bias of near-2:1 systems toward slightly wider separations is then a consequence of this asymmetry in the tidal evolution of planets at longer periods. With the current separation of planet b, a hypothetical inner planet in a 1:2 MMR would have a radius of only twice the host star's radius, which is $1.59\,M_{\sun}$.} The alternative proposed by \citet{weinberg} has its own set of problems. If a star slowly evolving toward the subgiant state suddenly experiences a transformation when the excited gravity waves begin to break in the radiative core increasing the tidal quality by orders of magnitude, all hot Jupiters should perish by merging with the star soon after that event. This prediction is in conflict with the observation that about 0.49\% of low-luminosity red giants have giant planets with orbital periods shorter than 10 days, which is nominally higher than the rate for solar-type dwarfs \citep{2019AJ....158..227G}.

 \section*{Acknowledgments}

 The authors are grateful to Pavel B. Ivanov for a valuable consultation on dissipation in stars.


\bibliography{references}
\bibliographystyle{plainnat}

 \end{document}